\documentclass[fontsize=11pt,twoside=semi,numbers=endperiod]{scrartcl}
\KOMAoptions{headinclude=true,footinclude=false,index=toc}
\usepackage[paperwidth=150mm,paperheight=179.35mm,width=110mm,totalheight=165.35mm,includehead,top=4mm,inner=20mm]{geometry} 

\usepackage{amsmath}
\usepackage{amssymb}

\usepackage{lmodern} 

\usepackage[ansinew]{inputenc}
\usepackage[T1]{fontenc}
\usepackage[english,ngerman]{babel}
\usepackage{textcomp} 
\usepackage[pdftex]{color,graphicx} %
\usepackage{bbm}
\usepackage{cite} 
\usepackage{ellipsis} 
\usepackage{microtype} 
\usepackage{scrpage2}
\usepackage[%
	pdftex,
	colorlinks,citecolor=blue,filecolor=blue,linkcolor=blue,urlcolor=blue,%
	bookmarks=true,%
	bookmarksopen=false,%
	bookmarksnumbered=true,%
	pdfpagemode=UseNone,
    pdfstartview={XYZ null null 1.00},
	pdffitwindow=true,
    pdfview={XYZ null null null},
    pdfpagelayout=SinglePage,
    pdfdisplaydoctitle=true, 
    plainpages=false,
    pdfpagelabels=true,
	pdfsubject={interpretation of quantum theory},
	pdfkeywords={Quantum Bayesianism, QBism, non-locality, interpretation of quantum theory},
	pdftitle={QBism, Quantum Nonlocality, and the Objective Paradox},
    pdfauthor={Astrophysical Institute Neunhof}]
	{hyperref}

\setkomafont{disposition}{\rmfamily} 
\addtokomafont{section}{\large\bfseries} 
\addtokomafont{subsection}{\normalsize\bfseries} 
\addtokomafont{pagenumber}{\LARGE}
\addtokomafont{caption}{\small}

\deffootnote[1em]{1em}{1em}{\textsuperscript{\thefootnotemark}\ }

\newcommand{\ggie}{\mbox{i.\,e.}\ }

\newcommand{\bz}[1]{\nolinebreak\hspace{0em}\nolinebreak{}#1\hspace{0em}}
\newcommand{\al}{\iflanguage{english}{``\nolinebreak\hspace{0em}}{\glqq\nolinebreak\hspace{0.1em}}\nolinebreak}
\newcommand{\ar}{\nolinebreak\iflanguage{english}{\hspace{0em}\nolinebreak{}''\ }{\hspace{-0.45em}\nolinebreak\grqq\thickspace}}
\newcommand{\arp}{\nolinebreak\iflanguage{english}{\hspace{0em}\nolinebreak{}''}{\grqq}}

\clearscrheadfoot 
\rohead[\vspace{-2mm}\par\pagemark ]{\vspace{-2mm}\par\pagemark}
\lehead[\vspace{-2mm}\par\pagemark ]{\vspace{-2mm}\par\pagemark}
\rehead{ }
\lohead{ }

\graphicspath{{Abbildungen/}}

\begin{document}
\selectlanguage{english}
\thispagestyle{empty}
\raggedbottom  
\vspace*{2mm}
\centerline{\bfseries\Large QBism, Quantum Nonlocality, and}
\vspace*{2mm}
\centerline{\bfseries\Large the Objective Paradox}
\vspace*{3mm}
\centerline{{\bfseries Gerold Gr\"{u}ndler}\,\footnote{\href{mailto:gerold.gruendler@astrophys-neunhof.de}{email:\ gerold.gruendler@astrophys-neunhof.de}}}
\vspace*{1mm}
\centerline{\small Astrophysical Institute Neunhof, N\"{u}rnberg, Germany} 
\vspace*{7mm}
\noindent\parbox{\textwidth}{\small\hspace{1em}The Quantum-Bayesian interpretation of quantum theory claims to eliminate the question of quantum nonlocality. This claim is not justified, because the question of non-locality does not arise due to any interpretation of quantum theory, but due to objective experimental facts. We define the notion ``objective paradox'' and explain, comparing QBism and the Copenhagen interpretation, how avoidance of any paradox results into poor explanatory power of an interpretation, if there actually exists an objective paradox.} 
\vspace*{5mm}

\section{Overview}
Quantum Bayesianism\!\cite{Fuchs:qbism1,Fuchs:qbismlocality}, or for short QBism, is an interpretation of quantum theory, which claims to \al remove the paradoxes, conundra, and pseudo\bz{-}problems that have plagued quantum foundations for the past nine decades\arp\!\cite{Fuchs:qbismlocality}, and in particular \al eliminates `quantum nonlocality'.\arp\!\cite{Fuchs:qbismlocality} The latter claim is challenged in section\:\ref{sec:nonlocality}\,. In section\:\ref{sec:explpower} the discussion is extended to a general consideration regarding the (relatively poor) explanatory power of QBism. In section\:\ref{sec:objparadox} the notion \al objective paradox\ar  is defined. We point out, how the anxious avoidance of any paradox --- even an objective paradox --- causes QBism's deficit of explanatory power. 

\section{Quantum Nonlocality}\label{sec:nonlocality}
We consider the gedanken\bz{-}experiment described by Einstein, Podolsky,  and Rosen\!\cite{Einstein:EPR} in the simplified variant proposed by Bohm\!\cite{Bohm:quantmech}: An instable system is prepared in a singlet state. It decays into two fragments with \mbox{$\text{spin}=1/2$} each. At position $x_A$, Alice measures by means of a Stern\bz{-}Gerlach magnet the spin of one of the fragments along the $z$\bz{-}axis. At position $x_B$, Bob measures by means of another Stern\bz{-}Gerlach magnet the spin of the other fragment along the $z$\bz{-}axis. We assume that the measurements at $x_A$ and $x_B$ are space\bz{-}like separated, such that Alice and Bob can not get any informations on the result of the other, before they have completed their own measurement. In the sequel we will discuss the experiment in a reference frame, in which Alice has completed her measurement, before Bob starts his measurement. This arbitrary choice of reference system is of no relevance regarding the question of nonlocality. 

Based on their knowledge of the experimental setup, Alice and Bob both describe the spin state of the singlet system, as long as they have not yet made their measurements, by the state function\pagebreak[1] 
\begin{align}
\sqrt{\frac{1}{2}}\,\Big(\, |\downarrow\,\rangle _A\otimes |\uparrow\,\rangle _B+|\uparrow\,\rangle _A\otimes |\downarrow\,\rangle _B\,\Big)\ .\label{kjsjngnhsfd} 
\end{align} 

Depending on the result $\uparrow _{A}$ or $\downarrow _{\,A}$ of her measurement, Alice assigns a state function to her own fragment 
\begin{subequations}\label{jmkdnmgfii}\begin{align}
\uparrow _{A}\ &\longrightarrow\ |\uparrow\,\rangle _A
&\downarrow _{\,A}\ &\longrightarrow\ |\downarrow\,\rangle _A\ , 
\end{align} 
and --- due to application of \eqref{kjsjngnhsfd} --- she assigns at the same time a state function to Bob's fragment: 
\begin{align}
\uparrow _{A}\ &\longrightarrow\ |\downarrow\,\rangle _B
&\downarrow _{\,A}\ &\longrightarrow\ |\uparrow\,\rangle _B\label{jmkdnmgfiib}
\end{align}\end{subequations} 
The state assignment \eqref{jmkdnmgfiib} to Bob's fragment just matches Alice's conditional probabilistic assumptions 
\begin{align}
P(\,\uparrow _{B}|\uparrow _{A})&=0&P(\,\downarrow _{\,B}|\uparrow _{A})&=1\notag\\ 
P(\,\uparrow _{B}|\downarrow _{\,A})&=1&P(\,\downarrow _{\,B}|\downarrow _{\,A})&=0 
\end{align} 
on Bob's result, which again are based on her knowledge of the fact that the primary system has been prepared in a singlet state. 

To EPR\!\cite{Einstein:EPR}, Alice's state assignment \eqref{jmkdnmgfiib} to Bob's fragment seemed a severe problem. They argued: If Alice can predict the outcome of Bob's measurement with certainty ($P=1$), then there must exist an element of reality (something `out there'), which causes Bob's measurement result. But according to quantum theory, only the overall singlet spin function \eqref{kjsjngnhsfd} exists as long as neither Alice nor Bob have started their measurements, while specific spin states of the fragments do not yet exist. Only when the fragments arrive at the Stern\bz{-}Gerlach magnets --- but not earlier! --- , then Nature decides for either $\uparrow _A$ and $\downarrow _{\,B}$, or for $\downarrow _{\,A}$ and $\uparrow _B$\,. 

Alice and Bob can not use the correlation of their measurement results for exchange of informations with superluminal speed. Thus with regard to Alice and Bob, special relativity theory is not violated. But Nature herself has the information on the spin settings of the fragments available over space\bz{-}like distance, with no time delay at all, as proved by the perfect correlation of Alice's and Bob's observations. One is tempted to say that Nature, when setting the spins in the moment of measurement, seems to be exempted from the restrictions of special relativity theory, and act non\bz{-}local. 

How does the Copenhagen interpretation explain the correlations? In his 1955 Gifford lecture\!\cite{Heisenberg:CopInt} on \al The Copenhagen Interpretation of Quantum Theory\arp , Heisenberg explicates his interpretation of the state function: 
\begin{addmargin}{1.5em}
\al The probability function combines objective and subjective elements. It contains statements about possibilities or better tendencies (`potentia' in Aristotelian philosophy), and these statements are completely objective, they do not depend on any observer; and it contains statements about our knowledge\footnote{Heisenberg uses the term \al knowledge\ar  for the subjective element represented by the state function. Timpson\!\cite[sec.\,2.3]{Timpson:qbism} emphasizes, that notions like \al knowledge\ar  or \al information\ar  have an objective character, as they can be objectively right or wrong. On the other hand, notions like \al assumption\ar  or \al believe\ar  denote something truly subjective. We probably understand Heisenberg correctly if we assume that he wanted to characterize the subjective element represented by the state function as \al believe\ar  of the physicist, but not as \al knowledge\arp , in contrast to the objective element represented by the state function (\ggie  the Aristotelian tendencies).} of the system, which of course are subjective in so far as they may be different for different observers. [\,\dots{}] we may say that the transition from the `possible' to the `actual' takes place as soon as the interaction of the object with the measuring device, and thereby with the rest of the world\footnote{Heisenberg was well aware of the importance of decoherence for the measurement process, long before Zeh and Zurek started to consider the issue in the seventies. If in doubt, read Heisenberg's 1955 article\!\cite{Heisenberg:CopInt}.}, has come into play; it is not connected with the act of registration of the result by the mind of the observer.\footnote{Note that Heisenberg here clearly addresses and removes the issue of \al Wigner's friend\arp , many years before Wigner\!\cite{Wigner:Freund} invented the alleged problem. QBism as well removes the issue, but in a basically different manner: In Heisenberg's point of view, the measurement result comes into being as soon as the instrument (which according to the Copenhagen interpretation must coercively be described by the methods and notions of classical physics) has registered the result, \ggie  already before the friend reads the result from the display, and a fortiori before he communicates the result to Wigner. In the QBism point of view, Wigner may describe the primary quantum object, the instrument, and the friend as an entangled quantum system, and the measurement result comes into being for Wigner only in the moment, when he receives his friend's report.} The discontinuous change in the probability function, however, takes place with the act of registration, because it is the discontinuous change of our knowledge in the instant of registration that has its image in the discontinuous change of the probability function.\ar
\end{addmargin}
The Aristotelian tendencies are not hard facts, but anyway they are something `out there', not merely assumptions or believes of a physicist. When Alice does her measurement, then \al the transition from the `possible' to the `actual' takes place\arp , and the Aristotelian tendencies `out there' are changed from \eqref{kjsjngnhsfd} to \eqref{jmkdnmgfii}. This of course does mean, that Alice due to her measurement triggers an objective change at Bob's place, with no time retardation. The measurement on the entangled system \eqref{kjsjngnhsfd} clearly has a non\bz{-}local character in Heisenberg's point of view. 

Bohr, on the other hand, never made (to my best knowledge) any onthological statement on the quantum\bz{-}theoretical state function.\footnote{The often quoted sentence \al There is no quantum world. There is only an abstract quantum physical description. It is wrong to think that the task of physics is to find out how nature \emph{is}. Physics concerns what we can \emph{say} about nature.\arp , reported by Petersen\!\!\cite{Petersen:BohrPhilos}, is strongly misleading, if taken out of context. In that article, Petersen explains: \al Traditional philosophy has accustomed us to regard language as something secondary and reality as something primary. Bohr considered this attitude toward the relation between language and reality inappropriate. When one said to him that it must be reality which, so to speak, lies beneath language, and of which language is a picture, he would reply, `We are suspended in language in such a way that we cannot say what is up and what is down. The word reality is also a word, a word which we must learn to use correctly.' Bohr was not puzzled by ontological problems or by questions as to how concepts are related to reality. Such questions seemed sterile to him.\ar  

Bohr always insisted that the quantum objects and the measuring devices applied for their observation together constitute the individual quantum pheonomena. Only in their entirety can quantum phenomena reasonably be discussed by human beings. Bohr never answered with yes or no the question, whether (only fantasized but not observed) quantum objects are \al real\ar  \emph{without} the (classical) measurement instruments, which are needed for their observation. Any answer to that question seemed pointless to Bohr, and a misuse of language.} He neither objected nor approved Heisenberg's partial identification of the state function with objective Aristotelian tendencies `out there' (the potentia of Aristotle's philosophy); he simply --- and obviously very deliberately --- was silent about this question. But he always emphasized the wholeness of quantum phenomena. In his reply\!\cite{Bohr:EPR} to EPR, Bohr spoke of phenomena \al where we have to do with a feature of \emph{individuality} completely foreign to classical physics.\ar  He emphasized the word individuality by italics, and he left no doubt that he meant this notion literally.\footnote{(latin) individual\,=\,not divisible} Bohr was convinced that the measuring instruments, which physicists apply for the observation of a quantum object, must be considered an integral, not separable part of the individual quantum phenomenon. In the EPR\bz{-}gedankenexperiment discussed above, the individual quantum phenomenon extends from Alice's Stern\bz{-}Gerlach magnet to Bob's Stern\bz{-}Gerlach magnet. When Alice does a measurement, then the \emph{whole} individual quantum phenomenon is affected. In Bohr's eyes, it would not be correct to say that Alice's measurement affects only her fragment. There only exists the one indivisible system \eqref{kjsjngnhsfd}, onto which Alice's instrument works, and only as an effect of her completed measurement we can reasonably speak of two new quantum systems \eqref{jmkdnmgfii}, which have been created by Alice's measurement out of the primary quantum system \eqref{kjsjngnhsfd}. 

Thus, while their wordings are different, Heisenberg and Bohr agree on the non\bz{-}local character of quantum phenomena, and thereby offer an explanation for the EPR\bz{-}correlations. This explanation may not please everybody, but at least it is an explanation, and it is a clear (affirmative) answer to the question of quantum nonlocality. 

To Einstein, Podolski, and Rosen, on the other hand,  the assumption of nonlocality seemed unacceptable. From the perfect correlations, they concluded that the fragments actually must have well\bz{-}defined spin states $\uparrow $ or $\downarrow $ along the $z$\bz{-}axis all the time, independent of any measurement. They deemed quantum theory incomplete, \ggie  they assumed that \eqref{kjsjngnhsfd}, while being correct, must be completed by insertion of `hidden variables', which reflect the spin states of the fragments, and determine the results of Alice's and Bob's measurements. We do not need to accept this explanation, but at least it is an explanation, and it is a clear (negative) answer to the question of quantum nonlocality. 

QBism, however, works around the question of nonlocality due to a quite strange and radical measure: This interpretation postulates that the one and only purpose of quantum theory is to help an agent to organize and optimize his\bz{/}her personal believes regarding his\bz{/}her future experiences, which of course will happen at his\bz{/}her future place, but not somewhere else. Fuchs\,et.\,al.\!\cite{Fuchs:qbismlocality} explain: 
\begin{addmargin}{1.5em}
\al QBist quantum mechanics is local because its entire purpose is to enable any single agent to organize her own degrees of belief about the contents of her own personal experience. No agent can move faster than light: the space\bz{-}time trajectory of any agent is necessarily timelike. [\dots ] Quantum mechanics, in the QBist interpretation, cannot assign correlations, spooky or otherwise, to space\bz{-}like separated events, since they cannot be experienced by any single agent. Quantum mechanics is thus explicitly local in the QBist interpretation. And that’s all there is to it.\ar 
\end{addmargin}
Now, EPR\bz{-}correlations exist not only in gedanken\bz{-}experiments. They have been experimentally confirmed, see for example \cite{Hensen:Belltest,Giustina:Belltest,Shalm:Belltest}. EPR raised a reasonable question, when they asked how Nature brings about the perfect correlations, whether Nature really is acting non\bz{-}local over space\bz{-}like distances, or whether there actually exist hidden variables, or whether there is any further possible explanation. 

No interpretation of quantum theory is obliged to offer an explanation for the experimentally confirmed correlations. It's perfectly fine, if \al no comment!\ar  is QBism's only answer to the question of nonlocality. But it is not sensible to say that with QBism \al the issue of nonlocality simply does not arise\arp\!\cite{Fuchs:qbismlocality}. It \emph{does} arise, independent of any interpretation of quantum theory, due to hard experimental facts.

\section{Explanatory Power}\label{sec:explpower}
The QBist point of view, that Alice's assignment of state functions \eqref{jmkdnmgfiib} exists only for Alice at her place, but does not affect anything at Bob's place, does merely say what does \emph{not} cause the perfect correlation. When Alice is using quantum theory, to assign the state function \eqref{jmkdnmgfiib} to Bob's fragment, then she makes implicit use of the non\bz{-}local character of quantum theory. Of course she may shrug shoulders and never ask what is behind the baffling power of quantum theory, to supply her locally with believes regarding the results of space\bz{-}like distant measurements, which mysteriously turn out true with no exception, and help her to win every bet. Paraphrasing Caves, Fuchs, and Schack\!\cite[sec.\,VI,\,par.\,3--5]{Fuchs:qbism2}, this is the statement of QBism: 
\begin{addmargin}{1.5em}
Doesn’t the perfect correlation demand an explanation independent of Alice’s belief?

Alice has put together all her experience, prior beliefs, previous measurement outcomes, her knowledge of physics and in particular quantum theory, all to predict the perfect correlation of her results with Bob's. Why would she want any further explanation? What could be added to her belief of certainty? She has consulted the world \emph{in every way she can} to reach this belief; the world offers no further stamp of approval for her belief beyond all the factors that she has already considered.
\end{addmargin}
These sentences are indicating a misunderstanding. Asking \al Why would she want any further explanation?\arp , Caves, Fuchs, and Schack deny that QBism actually has deprived Alice of \emph{any} explanation for the perfect correlations. The belief of certainty, and her unrestricted trust in the correctness and reliability of the abstract mathematical machinery of quantum theory can not replace anything which could rightly be named an \al explanation\arp . 

All reasonable interpretations of quantum theory (QBism no doubt belongs to this group) are identical with regard to the experimentally verifiable\bz{/}falsifiable consequences which can be derived from them. Thus reasonable interpretations differ not by being right or wrong (they all are right), they merely differ in the metaphysics, \ggie  in the intelligible pictures (positivists would say in the distracting illusions) they are offering to give us a framework in which we can sort our experiences. The coherence of this metaphysical framework then can give us the \emph{feeling to understand} the phenomena, which is much more than merely being able to compute and correctly predict future measurement results, and thereby sorting and optimizing one's personal believes regarding one's future personal experiences.  

Timpson\!\cite[sec.\,4.2]{Timpson:qbism} points out a general \al explanatory deficit problem\ar  of QBism, and presents besides many other examples the question, why some solids are good electrical conductors, while others are insulators. Independent of any interpretation, we can apply the mathematical machinery of quantum theory, and find out whether the Fermi surface is within a partially filled conduction band (then this solid is a conductor) or between and far\bz{-}off two conduction bands (then the solid is an insulator). If we follow the QBism interpretation, then the mathematical result helps us to update our personal believes and informs us, which bets we should accept, and which bets we better should reject, `and that's all there is to it.' With other interpretations, which consider the state function as representing something `out there', we get a much richer picture: We see electrons, which can (or can not) due to an externally applied voltage and\bz{/}or interactions with phonons be excited into a free energy level, and then move almost unimpeded through the solid. 

No such pictures exist with QBism, because in that interpretation the state function does not represent anything `out there'. This deficit of plausible pictures of course does not affect the predictive power with regard to future measurement results. Thus it is a matter of taste, whether we, following Timpson\!\cite{Timpson:qbism}, complain of QBism's lack of explanatory power, or whether we rather appreciate that QBism spares us much of the metaphysical ballast of other interpretations. 

Note by the way that the question, whether some field like the $\psi $\bz{-}field of quantum theory should be reified and considered to represent something `out there', does not only turn up in the interpretation of quantum theory. Consider for example Maxwell's electromagnetic field. Does it \al exist\arp ? It is created due to the dynamics of charged particles, and the one and only method to observe it is due to the observation of charged test particles. Thus the electromagnetic field may rightly be considered to be nothing than an abstract mathematical formalism, which helps us to organize our believes regarding the question, how the dynamics of some charged particles will impact the dynamics of some other charged particles. Still the mental picture of an electromagnetic field really existing `out there' is of such convincing explanatory power, and such a valuable help to develop our intuition about electromagnetic interactions, that many of us would strongly hesitate to give it up. 

E.\,T.\,Jaynes\!\cite{Jaynes:omelette} once described the formalism of quantum theory as\vspace{-2ex} 
\begin{addmargin}{1.5em}
\al a peculiar mixture describing in part realities of Nature, in part incomplete human information about Nature --- all scrambled up by Heisenberg and Bohr into an omelette that nobody has seen how to unscramble. Yet we think that the unscrambling is a prerequisite for any further advance in basic physical theory. For, if we cannot separate the subjective and objective aspects of the formalism, we cannot know what we are talking about\arp .\end{addmargin}
Caves, Fuchs, and Schack may rightly claim that they unscrambled the egg, at least with regard to the interpretation of the state function. They purged it from all objective content (Heisenberg's Aristotelian tendencies out there, Bohr's objective individuality of quantum phenomena), and kept nothing but the subjective believes of an agent. But this success comes at a high price: At the same time, they skipped a large part of the explanatory metaphysical power of the Copenhagen interpretation, without replacing it by anything better. 

What is so bad with the inseparable mixture of objective facts and subjective believes in quantum theory? We acquired our cognitive capabilities during millions of years of evolution. Hence it is no surprise that these capabilities don't fit to a type of phenomena (\ggie  quantum phenomena) which human beings first time encountered in the twentieth century. The `paradoxes, conundra, and pseudo\bz{-}problems' turning up in the discussion of quantum phenomena are not caused by any interpretation of quantum theory. Instead they are caused by the mismatch between objective facts `out there' and our human type of cognition, which is not prepared to cope with quantum phenomena. 

We have a theory, which perfectly matches our cognitive capabilities: Classical physics. A fictitious theory, which would perfectly match the objective world `out there' would be completely useless, because no human being would be able to understand it. Quantum theory can only be useful, if it spans the gap between the objective world and our cognitive capabilities. To meet this demand, quantum theory must necessarily combine elements of objective aspects of the world, and elements of the human type of thinking, like a bridge over a river must have bridge heads on both banks to serve it's purpose. Like no part of the bridge can be assigned to the left or to the right bank of the river, but each part is necessarily tied to both banks, the parts of quantum theory can not be separated into elements which belong to the objective world, and other elements which belong to the human observer. Instead quantum theory can only meet it's purpose, if  each of it's parts is reflecting both the objective reality `out there' and the cognitive capabilities and the type of thinking of the human observer. The inseparability of the subjective and objective aspects of the formalism is not a problem, but a necessary feature of quantum theory. 

Once we had the mathematical machinery of quantum theory, found by ingenious guessing, we next needed some interpretation, because we felt that a perfectly working mathematical machinery is not sufficient, to give us the feeling of true understanding. Bohr and Heisenberg did an excellent job, when they worked out the Copenhagen interpretation\!\cite{Heisenberg:CopInt}, to reconcile the objective world `out there' with the cognitive capabilities of human brains. Good ideas for further improvement are welcome. QBism, however, simply amputated the objective part from the interpretation of the state function (while keeping with no modification the full mathematical machinery with it's `scrambled objective and subjective elements'), and left us with a torso of marginal explanatory power, a bridge with only one head on one bank.

\section{Accepting/denying an objective paradox}\label{sec:objparadox}
In the well\bz{-}known experiment of Tonomura et.\,al.\cite{Tonomura:electroninterference}, single electrons go one by one through the biprism with two openings, and still the observation points of the electrons in the detector plane add up after many experimental runs to an interference pattern. 

The state function, which quantum theory assigns to each single electron, evolves through both openings of the biprism. If the state function is interpreted as representing something objectively existing out there and really moving through \emph{both} openings of the biprism in each single run of the experiment, then we get a mental picture which offers an explanation for the interference pattern. According to the Copenhagen interpretation, the electron may indeed be imagined as a wave moving through both openings. But at the same time the Copenhagen interpretation reminds us that the validity of the wave picture is limited by the complementary picture of a particle, and that it depends on our choice of the particular experimental arrangement, whether the wave\bz{-}like or the particle\bz{-}like character of the electron is elicited. Actually in the experiment of Tonomura et.\,al., the electron is elicited as a wave while moving through the biprism, but as a particle when being observed in the detector plane. We thus arrive in case of this experiment not at a consistent classical picture of wave \emph{or} particle, but at the complementary picture of wave \emph{and} particle. 

If we could have nice classical pictures, then hardly anybody would reject them. But if we can only have such strange and paradoxical complementary pictures, then some of us prefer to skip those dubious \al explanations\ar  completely, and constrain to well\bz{-}behaved Bayesian probabilities. While others think that strange explanations at least are better than no explanations. Its a matter of taste. Its \emph{really} nothing but a matter of taste, as either interpretation leads to identical experimentally testable consequences, and to identical expectation values for future measurements. 

QBism claims \al that it removes the paradoxes [\,\dots ] that have plagued quantum foundations for the past nine decades.\arp\!\cite{Fuchs:qbismlocality} The greek word paradox means, that something is incompatible with the human way of thinking. Is it really an advantage, if an interpretation of quantum theory can avoid any paradox? It may \emph{not} be an advantage, if there really should be an objective mismatch inbetween the reality of quantum phenomena and the cognitive capabilities of human beings, caused by the fact that quantum phenomena were irrelevant during the many millions of years of human evolution. Such mismatch could be named an \al objective paradox\arp , which can not be removed, because we can change the contents of our thinking, but we can not change our way of thinking. If such objective paradox really exists, shouldn't then an interpretation of quantum theory better acknowledge it, appropriately reflect it, and somehow arrange with it, instead of trying to shift it aside? 

\al The Copenhagen interpretation of quantum theory starts from a paradox.\ar  is the first sentence of Heisenberg's article\cite{Heisenberg:CopInt}. And some pages later he explains: 
\begin{addmargin}{1.5em}
\al There is no use in discussing what could be done if we were other beings than we are. At this point we have to realize, as von Weizs\"{a}cker has put it, that `Nature is earlier than man, but man is earlier than natural science'. The first part of the sentence justifies classical physics, with its ideal of complete objectivity. The second part tells us why we cannot escape the paradox of quantum theory, namely, the necessity of using the classical concepts.\ar 
\end{addmargin}
The Copenhagen interpretation acknowledges the existence of an objective paradox, and reflects it by the introduction of complementary explanatory pictures. QBism, on the other hand, targets --- and indeed accomplishes! --- an interpretation which is completely free of paradoxa. For this purpose it reduces the interpretation of the state function to representing the subjective assumptions of an agent, but not anything objectively existing `out there'. As an inevitable trade\bz{-}off, the denial of the objective paradox thereby deprives QBism of almost all explanatory power. 

The nonlocality of Nature, encoded in the objective individuality of quantum phenomena in Bohr's wording, or in objective Aristotelian tendencies in Heisenberg's wording, may seem quite strange to many of us. Indeed, these ideas \emph{really are very strange}, if considered in human brains, whose cognitive capabilities have been shaped by millions of years of interaction with an environment, which can appropriately be described in terms of classical physics.\footnote{How could Aristotle conceive the objective tendencies, even though he never observed a quantum phenomenon? Isn't this an indication, that the objective tendencies actually are not something `out there', but a subjective element of human cognition? Probably the best answer to this question is to note, that the request for a separation of objective and subjective elements is essentially nonsense, both with regard to the mathematical formalism and with regard to the interpretation of physical theories.} But is the wordless shrug of shoulders offered by QBism really better? Well, that's a matter of taste.\clearpage  

\flushleft{\interlinepenalty=100000\bibliography{../gg}} 
\end{document}